\documentclass[sort&compress,noshowpacs,preprintnumbers,amsmath,amssymb,aps,floatfix,preprint,showkeys]{revtex4}
 
\usepackage{graphicx}
\usepackage{epstopdf}
\usepackage{amsmath}
\usepackage{multirow}
\usepackage{times}
\usepackage{amssymb}
\usepackage{dcolumn}%
\usepackage{bm}%
\usepackage{hyperref}
\usepackage{upgreek}
\usepackage{wasysym}
\usepackage{verbatim}
\usepackage{blindtext}
\usepackage{xcolor}

\hypersetup
{
	colorlinks=true,
	linkcolor=blue,
	citecolor=blue
}

\newcommand{\tantala}{$\mbox{Ta}_{2}\mbox{O}_{5}$ }

\begin{document}

\title{Non-local cooperative atomic motions that govern dissipation in amorphous tantala unveiled by dynamical mechanical spectroscopy}


\author{F.\@ Puosi}
\email{francesco.puosi@pi.infn.it}
\affiliation{Istituto Nazionale di Fisica Nucleare, Largo B. Pontecorvo 3, I-56127 Pisa, Italy}
\affiliation{Dipartimento di Fisica ``Enrico Fermi'', 
Universit\`a di Pisa, Largo B.\@Pontecorvo 3, I-56127 Pisa, Italy}

\author{F.\@ Fidecaro}
\affiliation{Dipartimento di Fisica ``Enrico Fermi'', 
Universit\`a di Pisa, Largo B.\@Pontecorvo 3, I-56127 Pisa, Italy}
\affiliation{Istituto Nazionale di Fisica Nucleare, Largo B. Pontecorvo 3, I-56127 Pisa, Italy}

\author{S.\@ Capaccioli}
\affiliation{Dipartimento di Fisica ``Enrico Fermi'', 
Universit\`a di Pisa, Largo B.\@Pontecorvo 3, I-56127 Pisa, Italy}
\affiliation{Istituto Nazionale di Fisica Nucleare, Largo B. Pontecorvo 3, I-56127 Pisa, Italy}

\author{D.\@ Pisignano}
\affiliation{Dipartimento di Fisica ``Enrico Fermi'', 
Universit\`a di Pisa, Largo B.\@Pontecorvo 3, I-56127 Pisa, Italy}
\affiliation{Istituto Nazionale di Fisica Nucleare, Largo B. Pontecorvo 3, I-56127 Pisa, Italy}

\author{D.\@ Leporini}
\affiliation{Dipartimento di Fisica ``Enrico Fermi'', 
Universit\`a di Pisa, Largo B.\@Pontecorvo 3, I-56127 Pisa, Italy}
\affiliation{Istituto Nazionale di Fisica Nucleare, Largo B. Pontecorvo 3, I-56127 Pisa, Italy}

\begin{abstract}
 
The mechanisms governing mechanical dissipation in amorphous tantala are studied at microscopic scale via Molecular Dynamics simulations, namely by mechanical spectroscopy in a wide range of temperature and frequency. We find that dissipation is associated with irreversible atomic rearrangements with a sharp cooperative character, involving tens to hundreds of atoms arranged in spatially extended clusters of polyhedra. Remarkably, at low temperature we observe an excess of plastically rearranging oxygen atoms which correlates with the experimental peak in the macroscopic mechanical losses. A detailed structural analysis reveals preferential connections of the irreversibly rearranging polyhedra, corresponding to edge and face sharing. These results might lead to microscopically informed design rules for reducing mechanical losses in relevant materials for structural, optical, and sensing applications. 
\end{abstract}



\maketitle

\section{Introduction}

Mechanical dissipation in solids is a key topic in condensed matter, with very high scientific and technological significance. Contrary to crystals, where mechanical losses via energy dissipations are mainly due to lattice defects, dislocations and impurities, in disordered materials such as amorphous solids and glasses diverse phenomena can play a role, depending on temperature and frequency. For instance, restricting to temperatures above $\sim 10\,\mbox{K}$ where quantum tunneling is negligible and moving from THz to acoustic frequencies, the dominant contributions to dissipation span from Rayleigh scattering \cite{VacherCourtensPRB05_I,CourtensRuffleMonacoRayleighScatteringPRL06,BaldiMonacoRutaRayleighPRL10,RuoccoSchirmacherNatCom13,RuffleAttenuationPowerLawSilicaPRB17}, to Akhiezer damping \cite{NanomechanicalResonatorRourke2016,Maris1971,VacherCourtensPRB05_I,VacherCourtensPRB05_II,RuoccoSchirmacherNatCom13}, thermoelastic damping \cite{NanomechanicalResonatorRourke2016,Zener_1940,NowickBerryAnelasticRelax} and thermally activated relaxations \cite{PhillipsTLS_1987,VacherCourtensPRB05_II,HamdanJCP14}. 

Mechanical losses in materials can be estimated assuming a cyclic dissipation point of view, namely focusing on the energy loss during cyclic mechanical loading. In this way, considering  the loss angle $\phi$ between an imposed oscillatory strain and the corresponding stress in the material, the mechanical losses are obtained via the reciprocal quality factor, $Q^{-1}=\tan \phi$.  Dynamical mechanical spectroscopy (DMS) is widely used to study  liquids and glasses  experimentally \cite{CasaliniMechSpec13,CrespoReviewMechSpec2015,Rubinstein} as well as numerically through molecular dynamics (MD) simulations \cite{BarratMechSpecMD06,BarratMechSpectPRL19,BaschnagelMechSpecMolPhys15,DamartRodneyMechSpectrPRB17,ZacconeMechSpectSoftMatter18}. 

In a recent paper \cite{Puosi_PhysRevRes2019}, we applied DMS by simulations to investigate the mechanical losses in amorphous tantala ($\mbox{Ta}_{2}\mbox{O}_{5}$) in a wide range of frequency and covering  room temperature as well as cryogenic regimes. The results from simulations remarkably agreed with the experimental data available for annealed amorphous films, especially in terms of the temperature dependence of the quality factor. 

Here, we extend this analysis and investigate the microscopic processes which are responsible for the mechanical losses in amorphous tantala. There have been a few numerical studies on oxide glasses \cite{HamdanJCP14,DamartRodneyMechSpectrPRB17,DamartPRB18}, and more specifically on tantala \cite{Trinastic_JCP13,TrinasticPRB16,Damart_JAP16}, aiming to clarify the atomic origin of energy dissipation.   The source of dissipation has been suggested in low-energy excitations involving small groups of atoms with temperature-dependent extension  and organization. Such excitations are commonly modeled in the framework of the two-level system (TLS) model \cite{Jackle_TLS_JNCS76,GilroyPhillipsPhilMag81}, i.e. as transitions between pairs of local  energy minima in the potential energy landscape (PEL). At high temperature,  above $T\sim10\,\mbox{K}$ where tunneling across the barrier becomes less efficient, transitions are mainly due to thermally-activated processes which may interact with mechanical excitations propagating in the solid. Within this approach, mechanical losses are indirectly estimated via a suitable exploration of the PEL and a discrete sampling of TLSs. 

At variance with TLS modeling, in this work we adopt an alternative approach to study dissipation at the microscopic level, building on mechanical spectroscopy. This allow us to directly identify the atomic rearrangements which govern dissipation in different temperature ranges and to establish a direct connection with the observed macroscopic behavior. 

\section{Model and methods}
\label{modmeth}

We carried out classical Molecular Dynamics simulations for amorphous tantala using LAMMPS software \cite{lammps}. Tantala is modeled via a modified van Beest, Kramer, and van Santen (BKS) potential \cite{BKS_PRL90} with an additional pseudo-covalent Morse term \cite{Trinastic_JCP13}. To speed up computations, we implemented Wolf truncation with a cut-off function as proposed in Ref. \onlinecite{Damart_JAP16}. Each simulation consists of $2520$ atoms, contained in a cubic box with periodic boundary conditions. Glassy samples were produced cooling high-temperature liquids. The tantala crystal is first equilibrated  at $300\,\mbox{K}$ and then rapidly heated to $5000\,\mbox{K}$. The liquid at $5000\,\mbox{K}$ is equilibrated for $50\,\mbox{ns}$ and then cooled down to   $0\,\mbox{K}$ at constant  rate in the NPT ensemble. During the quench, configurations at the temperatures of interest were collected, equilibrated again for  $50\,\mbox{ps}$ and finally energy-minimized. With this recipe, we generated $30$ independent samples, whose amorphous structure is confirmed after examining  their pair distribution function.

DMS was performed by imposing to the simulation box a  sinusoidal tensile strain $\epsilon_{ii}(t)=\epsilon_0\sin(\omega t)$ in the $i$-direction ($i=X, Y$ or $Z$), and measuring the corresponding tensile stress along the same direction, $\sigma_{ii}$. No deformation of the simulation box takes place in other directions. The results were averaged over all the three directions, and the frequency $f=\omega/2\pi$ was varied from $0.5\,\mbox{GHz}$ to $1\,\mbox{THz}$. We fixed the amplitude $\epsilon_0=0.01$, such that the deformation is in the {\it linear} elastic regime. A Berendsen thermostat was employed to maintain constant temperature conditions and dissipate the heat produced during the deformation. The frequency dependent quality factor $Q(f)$ is determined by the ratio $Q(f)=E'(f)/E''(f)$ where $E'(f)$ and $E''(f)$ are the storage and the dissipative parts of the dynamic elastic modulus.

\section{Results and discussion}
\label{resdisc}

\begin{figure}[t]
\begin{center}
\includegraphics[width=0.8\linewidth]{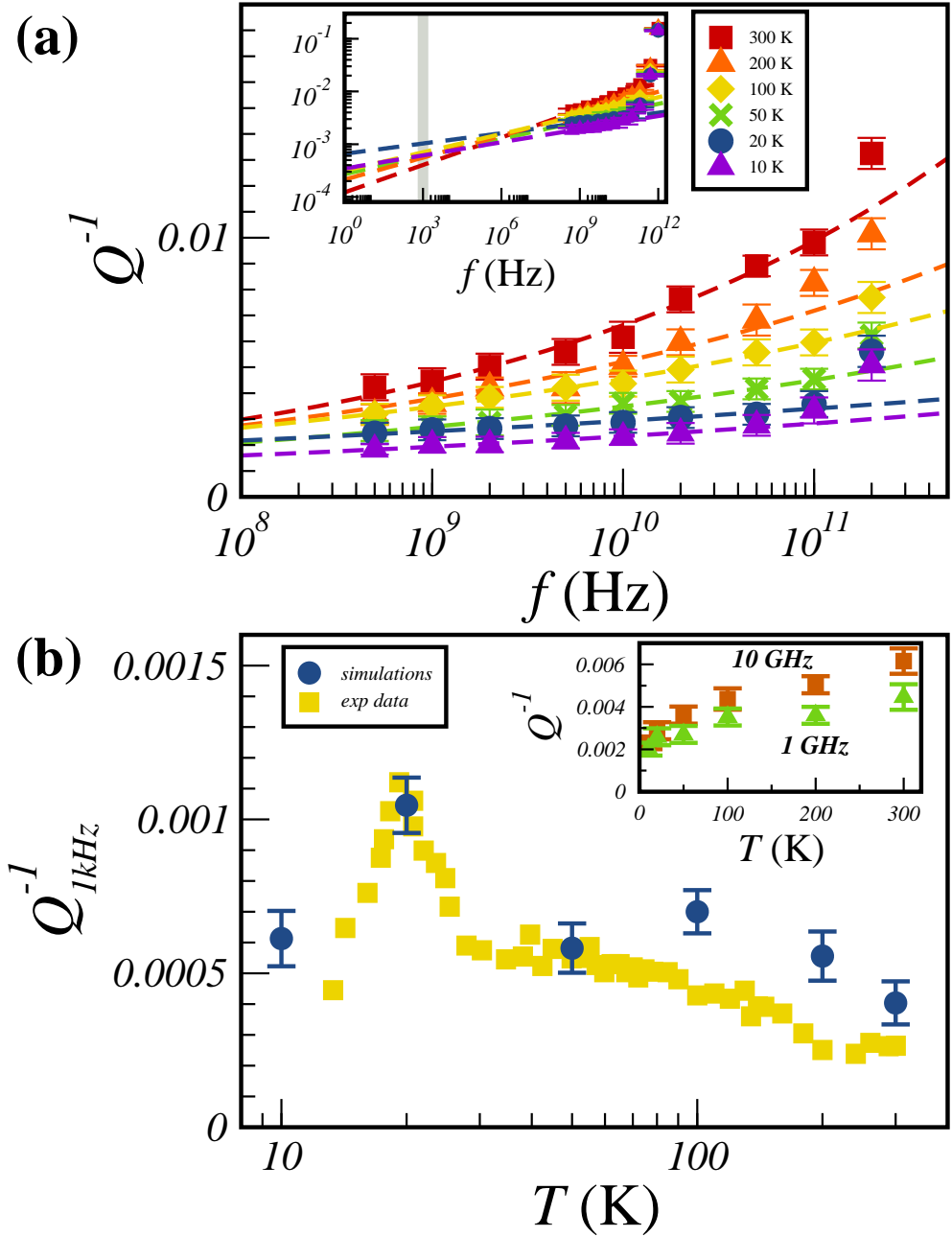}
\end{center}
\caption {Panel a): reciprocal quality factor $Q^{-1}$ as a function of frequency, $f$, for different sample temperature. The dashed lines are best-fit curves with a power-law relation. Inset:  low-frequency extrapolation of the power-law regime. Vertical line marks $f=1\,\mbox{kHz}$, which defines  $Q^{-1}_{1kHz}=Q^{-1}(f=1\,\mbox{kHz})$. Panel b):  temperature dependence of $Q^{-1}_{1kHz}$. For comparison, experimental data from Ref. \cite{d5} are also shown.  Inset: temperature dependence of $Q^{-1}$ at selected frequencies $1\,\mbox{GHz}$ and $10\,\mbox{GHz}$.}
\label{fig1}
\end{figure}

The frequency dependence of the reciprocal quality factor of \tantala glasses, as obtained by DMS analysis at different temperatures, is illustrated in Figure \ref{fig1}(a). A well-defined  power-law frequency dependence of $Q^{-1}$ is found in the GHz region, i.e., $Q^{-1}\sim f^{\alpha}$ where the exponent $\alpha$ ranges in $0.1-0.2$ with a non-monotonic temperature dependence \cite{Puosi_PhysRevRes2019}. Upon extrapolating the power-law behavior down to lower frequencies (inset of Fig. \ref{fig1}(a)), we estimate the value of  $Q^{-1}$ at $1\,\mbox{kHz}$ ($Q^{-1}_{1kHz}$),  which is of major interest  for many technologies in the broad fields of structural applications, photonics, interferometers, and sensors \cite{OptomechanicsRMP14,TangRoukeNatureNanotech07,BagciOpticalRadiowaveNature14}. In Figure  \ref{fig1}(b) we show  $Q^{-1}_{1kHz}$ as a function of temperature. Remarkably, $Q^{-1}_{1kHz}$ exhibits a peak at cryogenic temperature ($T\approx 20\,\mbox{K}$), in agreement with experimental findings in amorphous tantala films \cite{d5}.  We argue that this effect has a non-trivial origin: it emerges as the interplay between the $T$-dependence of mechanical losses in the high-frequency region (losses decreasing upon decreasing $T$, see the inset of Fig. \ref{fig1}(b)), and the underneath power-law behavior $Q^{-1}\sim f^{\alpha}$ (see Fig.\ref{fig1}(a)).  

\begin{figure}[t]
\begin{center}
\includegraphics[width=\linewidth]{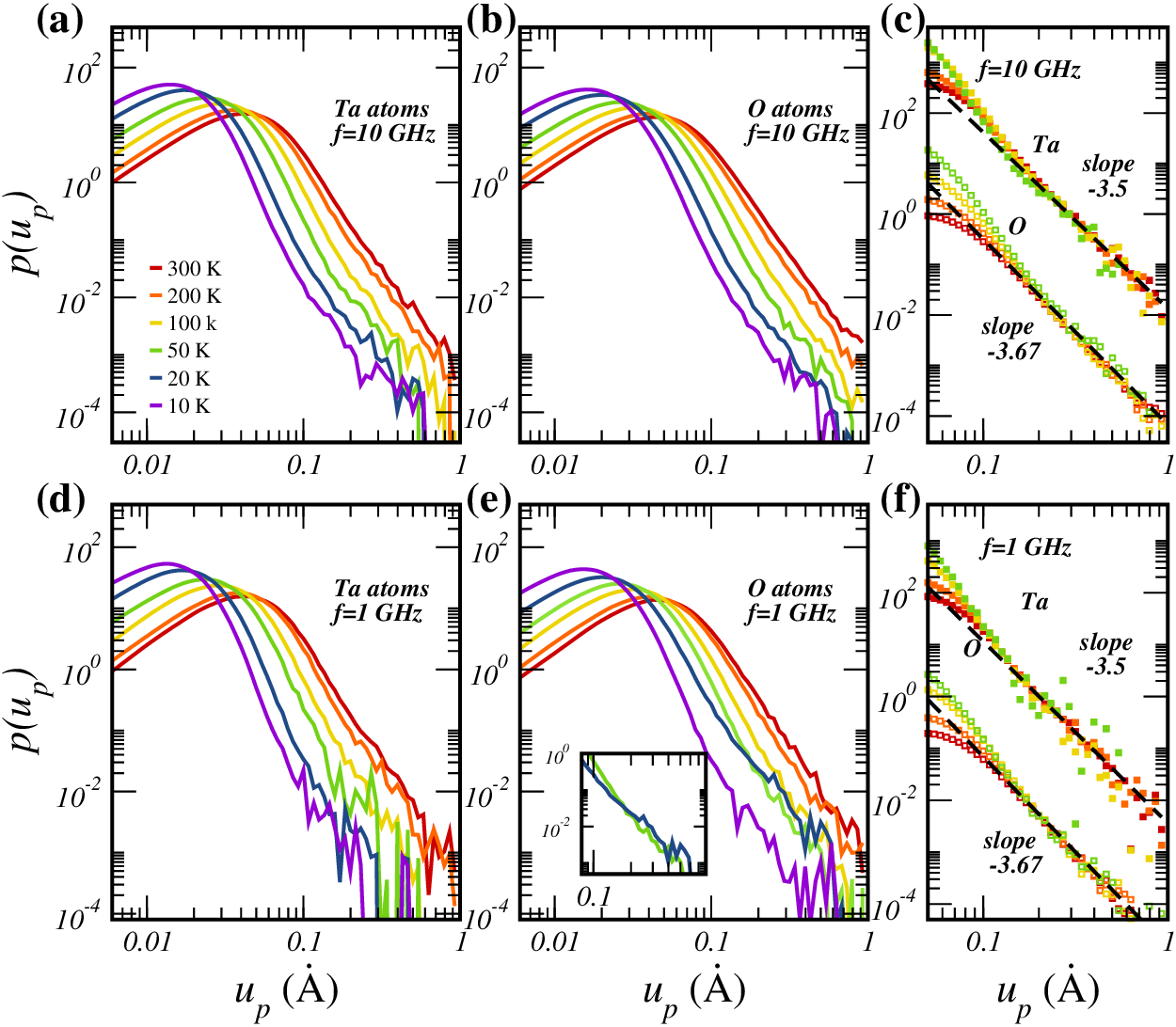}
\end{center}
\caption {Panel a),b),d) and e): distribution of the displacement $u_p$  for Ta and O atoms and for different frequencies, $10\,\mbox{GHz}$ (a,b) and $1\,\mbox{GHz}$ (d,e). Inset of panel e): magnification of the large-displacement tail of $p(u_p)$ for O atoms at frequency $1\,\mbox{GHz}$ and temperature $20\,\mbox{K}$ and $50\,\mbox{K}$. Panel c) and f): collapse of the tail of $p(u_p)$ via vertical rescaling. Dashed lines are best-fit curves with a power-law.   }
\label{fig2}
\end{figure}

Now we investigate the underlying atomic rearrangements that are responsible for the dissipation in amorphous $\mbox{Ta}_{2}\mbox{O}_{5}$. To this aim we consider the atomic displacement $u_p$ during a time interval $\Delta t$ corresponding to a deformation period $T_\omega$, starting from an unstrained configuration. In order to remove the effect of thermal vibrations, $u_p$  is calculated between minimum energy configurations, i.e. inherent structures (IS) \cite{Stillinger_Science1935}. 
We point out that the use of IS aims to expose more clearly the transitions between different configurations in the potential energy landscape of the system which do occur due to thermal activation and mechanical driving. 
In the following, we will refer to non-vanishing $u_p$ as ``plastic displacement''.  Figure \ref{fig2} shows the distribution  $p(u_p)$ for Ta and O atoms at selected frequencies, $1\,\mbox{GHz}$ and $10\,\mbox{GHz}$, and for different temperatures.  We note that for $T\geqslant 50 \,\mbox{K}$ $p(u_p)$ is nearly independent of the frequency for both Ta and O atoms. 
The peak of $p(u_p)$, which represents the most probable value of $u_p$, moves toward smaller values upon decreasing the temperature. Thus,  lower mechanical losses are associated with a reduced global atomic mobility.  The higher the temperature, the farther $p(u_p)$ extends. The tails of $p(u_p)$ exhibit a power-law behavior with exponents $-3.5$ for Ta and $-3.67$ for O, which seem not to depend on temperature and frequency (Fig. \ref{fig2}(c,f)). 

We notice that typical values of $u_p$ are fractions of interatomic distances. This could appear strange if compared to what is found in other systems,  e.g., metallic glasses, where $u_p$  approaches the interatomic distances \cite{Yu_PRB14}.  Still, the present  $u_p$  values are coherent with the magnitude of mechanical dissipation in the system:  small dissipation (inverse quality factor $Q^{-1}\lesssim10^-2$) correlates to  reduced atomic motions. In metallic systems  dissipation is ten to hundreds times stronger and accordingly atomic displacements  are also larger.

An interesting effect is observed for O atoms at the lowest frequency $1\,\mbox{GHz}$ (see Fig. \ref{fig2}(e), main panel and inset). For $T=20\,\mbox{K}$, the tail of $p(u_p)$ signals an excess of atoms with high mobility, i.e. a plastic displacement $u_p \geqslant 0.1\,\mbox{\AA}$.  We recall that at $20\,\mbox{K}$ low-frequency mechanical losses are maximum, as denoted by the  peak in $Q^{-1}_{1kHz}$. 

The scenario that we delineate builds on the actual definition of $u_p$ which identifies irreversible rearrangements over one deformation cycles. Reversible motions are also present. Recently Wang and coworkers \cite{Wang_AM2020} reported  the presence of a low temperature peak, at $T\sim 0.3 T_g$, in the relaxation spectrum of a model metallic glass. It originates from fast  relaxation processes with reversible character. From the previous analysis reversible motions seem to play a minor role in the present system. The source of this discrepancy could be due to the system details, oxide against metallic glass, and in particular to the presence in \tantala of oxygen atoms mainly contributing to dissipation. Differences in temperature and frequency/time ranges could also be relevant. Anyway, the role of reversible motions in dissipation is an interesting aspect which definitely deserves further attention. We plan to address it in future works.      

\begin{figure}[t]
\begin{center}
\includegraphics[width=\linewidth]{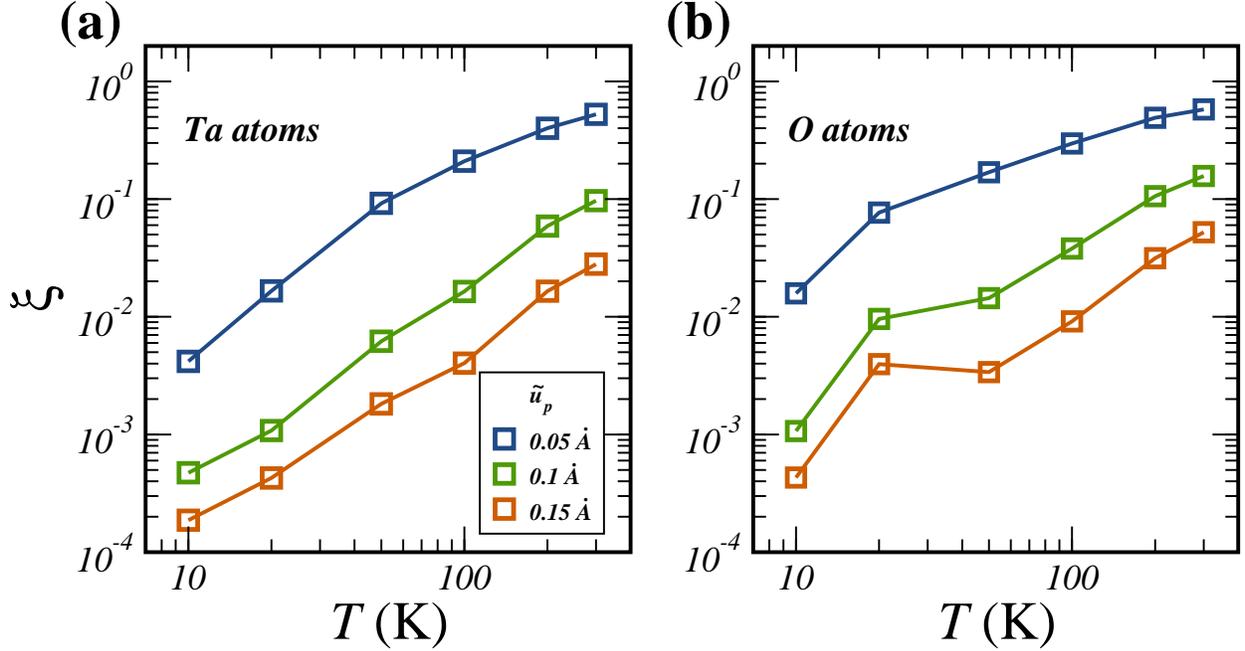}
\end{center}
\caption {Temperature dependence of the fraction of Ta atoms (panel a) and O atoms (panel b) with displacement $u_p$ exceeding a threshold value $\tilde{u}_p$.  }
\label{fig3}
\end{figure}

To better characterize the atomic motion that governs mechanical dissipation, we focus on the group of atoms with largest mobility. For sake of simplicity, we will restrict the following discussion to data obtained with a frequency $1\,\mbox{GHz}$.  The temperature dependence of the fraction $\xi$ of atoms with plastic displacement larger than a critical value  $\tilde{u}_p$ is shown in Figure \ref{fig3} for different choices of $\tilde{u}_p$. In the case of Ta atoms (panel a), $\xi$ behaves as one could expect, namely it decreases monotonically upon decreasing the temperature or increasing the threshold $\tilde{u}_p$. For O atoms (panel b), $\xi$ clearly highlights the excess of mobile atoms, observed in $p(u_p)$, which is signaled by a shoulder at $20\,\mbox{K}$ for $\tilde{u}_p=0.1\,\mbox{\AA}$, evolving to a peak if larger plastic displacements are considered. 

We next investigate the spatial organization of this fraction displaying significant mobility. It is well known that the structure of \tantala can be described as a network of Ta-centered polyhedra, mainly octahedra, with oxygen atoms at their corners \cite{Damart_JAP16,TrinasticPRB16,Prasai_PRL2019}. Two neighboring polyhedra can be classified according to how many O atoms are shared by the metal-metal pair: two Ta atoms sharing one, two or three O atoms are called corner-sharing (CS), edge-sharing (ES) or face-sharing (FS) polyhedra, respectively. On this basis we have performed a  detailed analysis of high mobility (HM) polyhedra. The following procedure is adopted to  define the set of relevant atoms. First we select the $5\% $ of atoms, both Ta and O, with largest plastic displacement $u_p$. Then we expand the selection by including all the other atoms which are also members of the polyhedra incorporating the initial HM atoms.  In this way we end up with a collection of tetrahedra, each containing at least one HM atom. 

\begin{figure}[t]
\begin{center}
\includegraphics[width=\linewidth]{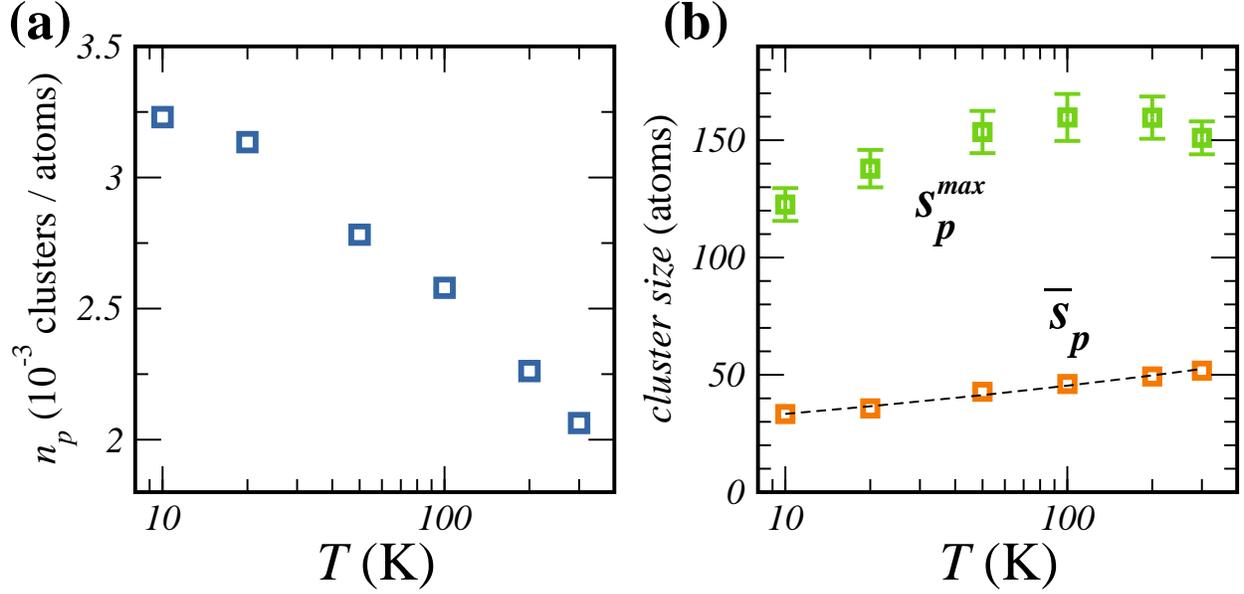}
\end{center}
\caption {Cluster analysis of HM polyhedra. Panel a): Temperature dependence of the mean density of HM clusters, $n_p$. Panel b): Temperature dependence of $\overline{s}$ and $s_p^{max}$ (the mean and largest cluster size, respectively), in number of atoms.   }
\label{fig4}
\end{figure}

\begin{figure}[t]
\begin{center}
\includegraphics[width=\linewidth]{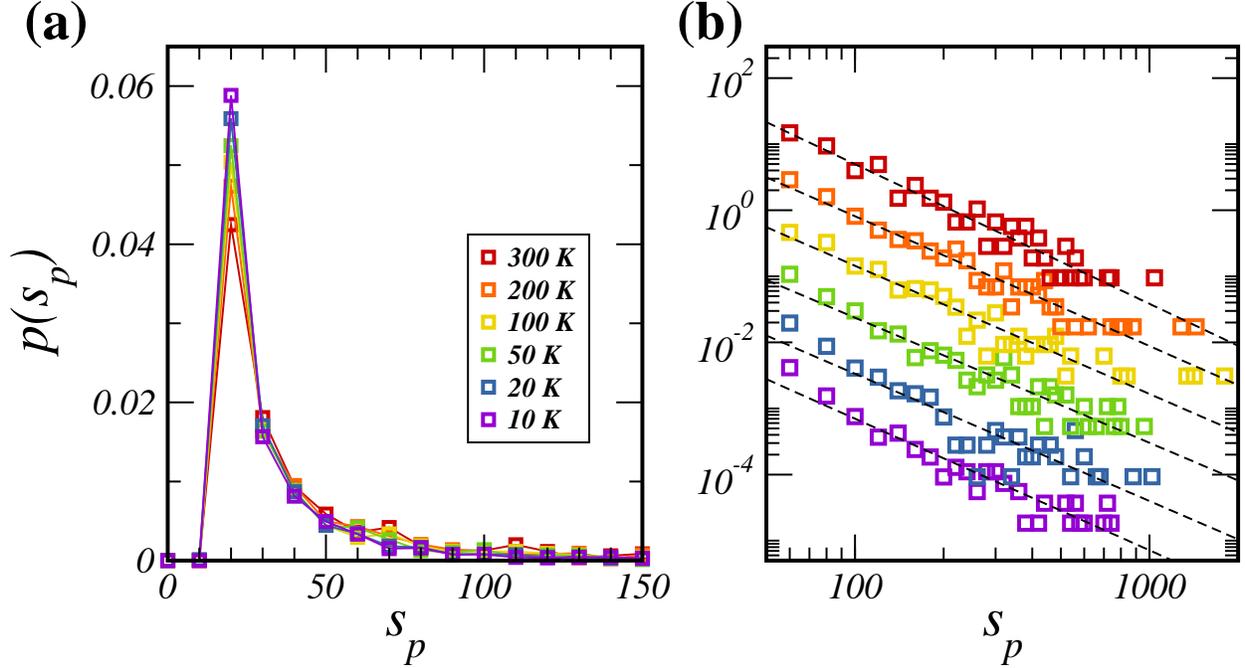}
\end{center}
\caption {Panel a): Distribution $p(s_p)$ of the population of HM clusters for different temperatures. Panel b): log-log plot of the tail of the distribution $p(s_p)$.  For clarity reasons, data are vertically shifted. Dashed lines are best-fit  power-law curves, with exponents $-2.11$, $-1.96$, $-1.94$, $-1.89$, $-1.93$ and $-1.99$ (listed from high to low temperature).   }
\label{fig5}
\end{figure}

The results of a cluster analysis of HM polyhedra are shown in Fig. \ref{fig4}. The density of cluster, $n_p$, defined as the number of HM clusters divided by the number of atoms in the system,  increases when moving from room to cryogenic temperatures (panel a).   A decrease upon cooling is observed in the mean cluster size $\overline{s}_p$ and the size of the largest cluster at a given time $s_p^{max}$, with the latter being virtually constant down to $50\,\mbox{K}$. These findings suggest that the dissipation mechanism changes with temperature, being dominated by few, large and irreversible atomic rearrangements at high temperature and by many and more localized events at cryogenic temperature.   
We also examine the distribution of cluster size, $p(s_p)$, in Figure \ref{fig5}. $p(s_p)$ features a peak around $s_p\sim20-30$ atoms whose height decrease upon increasing $T$,  which is also in agreement with the overall dissipation framework. The tail of the distribution follows a power law behavior (panel b) with an exponent $1.9\pm0.1$ which does not depend on temperature. 

Numerical simulations of DMS have been recently used to study the atomic rearrangements that govern internal friction in a model metallic glass \cite{Yu_PRB14}. The authors reported that ``faster atoms'' exhibit a strong tendency to aggregate and to form cluster-like structures whose sizes are distributed following a power law with exponent $1.6\pm0.2$  for the small to intermediate clusters (up to $\sim50$ atoms). Here we observe  for \tantala a power-law regime with a similar exponent which, however, involve larger clusters. 

\begin{figure}[t]
\begin{center}
\includegraphics[width=\linewidth]{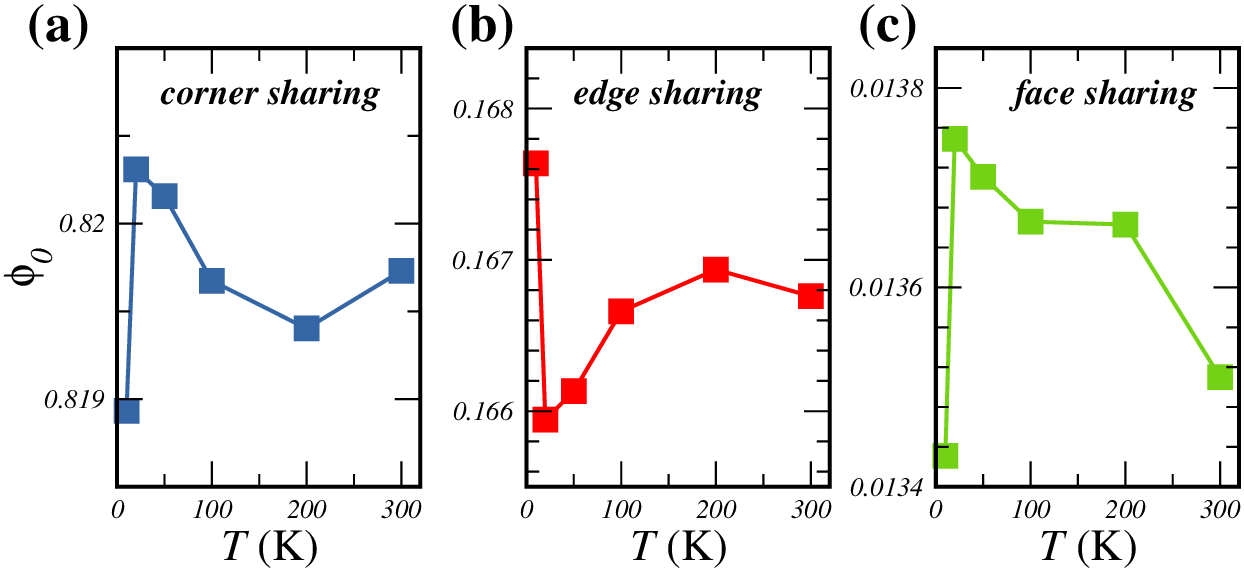}
\end{center}
\caption { Temperature evolution of the fraction $\phi_0$ of corner sharing (a), edge sharing (b) and face sharing (c) Ta-centered polyhedra.   }
\label{fig6}
\end{figure}

We also inspected the  connections between HM polyhedra.  As a preliminary step, in Figure \ref{fig6} we show the $T$-dependence of the fractions of CS, ES and FS polyhedra in quenched configurations at rest. In accord with previous results \cite{Prasai_PRL2019}, CS represents the main contribution, with a weight of more than $80\%$, followed by ES and FS links. Going more into the details of the comparison, in ref. \onlinecite{Prasai_PRL2019}, atomic modeling is adopted to interpret the effects of annealing tantala films at different temperature. Annealing is shown to increase the concentration of CS polyhedra and decrease those of ES and FS polyhedra, with the effect being enhanced upon increasing the annealing temperature.  We note that the specific fractions of CS, ES and FS polyhedra that we detect in our quenched configurations are compatible with those of annealed samples (at $T\approx600^\circ\mbox{C}$) analyzed in Ref. \onlinecite{Prasai_PRL2019}. This supports the previous findings \cite{Puosi_PhysRevRes2019} of a similarity in disordered structures created by different routes, quench cooling and deposition.   To conclude this analysis, we point out that the dependence on quenching temperature observed in Figure \ref{fig6} is rather weak,  with a less than $1\,\%$ variation in the investigated $T$ range, which makes it difficult to establish a correlation between the changes in the quenched structure and the behavior of  mechanical losses.

\begin{figure}[t]
\begin{center}
\includegraphics[width=\linewidth]{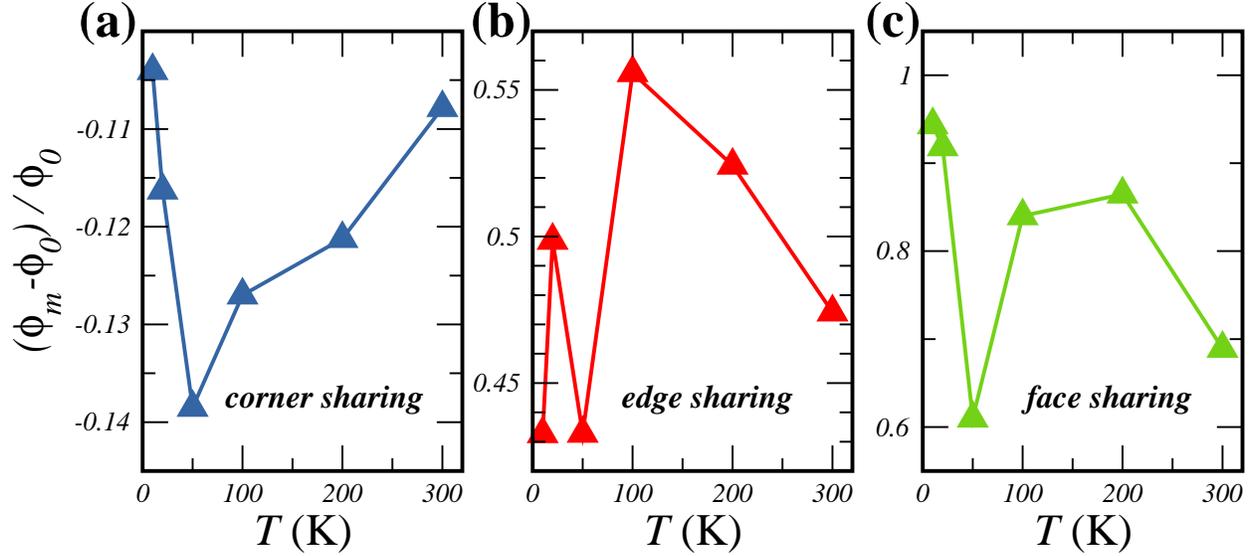}
\end{center}
\caption { Temperature evolution of the fraction $\phi_m$ of CS (a), ES (b) and FS (c)  HM Ta-centered polyhedra, relative to the corresponding global fraction   $\phi_0$.  }
\label{fig7}
\end{figure}

These results provide a reference for the study of polyhedra connections in HM clusters, whose main results are summarized in Figure \ref{fig7}. Here we focus on the temperature dependence of the relative variation of the polyhedra concentrations in HM clusters $\phi_m$, with respect to the populations of quenched states  $\phi_0$. We note that in HM clusters the populations of ES and FS polyhedra are enhanced, with deviations of $\sim50\%$ and $\sim80\%$, respectively, from the system populations. At the same time, CS polyhedra are less abundant in HM clusters, with a depletion by $\sim10\%$. Despite these variations, the hierarchy of populations is preserved, being $\sim0.72$, $\sim0.25$ and $\sim0.03$ the fractions of CS, ES and FS, respectively. 

This analysis points to a dissipation mechanism as due to HM atoms ascribed to ES and FS polyhedra. As we mentioned, in Ref. \onlinecite{Prasai_PRL2019} the reduction of mechanical losses via annealing at higher temperature  was associated with  an increase of CS and decrease of ES and FS polyhedra. Therefore, ES and FS are identified as the ``bad actors"  of the dissipation process in tantala, which is also the main conclusion of the analysis presented in Figure \ref{fig7}.

\section{Conclusions} 
\label{concl}

We have presented a detailed analysis of the microscopic processes which are responsible for mechanical losses in amorphous tantala. Building on our previous work  \cite{Puosi_PhysRevRes2019}, we adopted an approach based on MD simulations of mechanical spectroscopy. Dissipation is found to be governed by the fractions of atoms that perform irreversible plastic displacement in a time window corresponding to one deformation cycle. Remarkably, at low temperature we observe an excess of plastically rearranging oxygen atoms which correlates with the peak in the macroscopic mechanical loss (i.e., $Q^{-1}$). 
It is noteworthy to recall that, while plastic events are detected at GHz frequencies, the cryogenic dissipation peak occurs at much lower frequencies, in the kHz range. 
Even if the observed correlation may appear a specific feature of the investigated system, its relevance has a more general character. Indeed, it confirms the validity of the previously reported extrapolative approach, allowing simulations to be connected with experimentally-relevant frequency ranges \cite{Puosi_PhysRevRes2019}. Next natural step will consist to extend this analysis to other amorphous materials. As dissipation in amorphous \tantala shows striking similarities with other materials \cite{Ranganathan_JAP_2017}, like different oxides or metallic glasses, we are confident on the outcome of this method. 
%
%

We have also carried out a structural analysis of rearranging atoms. Spatially, plastic atoms exhibit a cooperative character, being arranged in clusters whose size can reach up to a few hundreds of atoms. Structurally, polyhedra comprising plastic atoms display an increased tendency to group via ES and FS connections, at the expense of CS links. This agrees with recent results in \tantala \cite{Prasai_PRL2019} associating a reduction in mechanical losses to an increase in CS polyhedra. 

Mechanical losses in amorphous oxides are critically important for nanosystems and gravitational wave detection \cite{GWReview2,GWReview1,GWReview3,GWReview4,Waggoner_JAP10,Battu_AEM2018,Granata_CQG2020}. We believe that a fundamental understanding of the atomic mechanisms at the base of dissipation in these materials could provide a guidance for the ongoing research activity which aims to a reduction of mechanical losses via the design of new materials or innovative manufacturing methods.  
 Indeed, if dissipation is demonstrated to originate from well identified ``sites'' like specific chemical species or local arrangements of atoms, a suitable design strategy could consist in  the reduction of their density or in the inhibition of their mobility via, for example, composition modification or doping. 

\begin{acknowledgments}
This project has received funding from the European Union's Horizon 2020 research and innovation programme under the Marie Sklodowska-Curie grant agreement No 754496. A generous grant of computing time from IT Center, University of Pisa and Dell${}^\circledR$ Italia is gratefully acknowledged.
\end{acknowledgments}

\bibliographystyle{elsarticle-num}
\bibliography{Virgo.bib}

\end{document}